\documentclass[useAMS]{mn2e}
\usepackage{amsmath}
\usepackage{times}
\usepackage{graphicx}

\title[X-ray behavior of XTE J1810-189]
{X-ray softening during the 2008 outburst of XTE J1810-189}

\author[Weng et al.]{Shan-Shan Weng$^{1,2,3}$\thanks{E-mail: wengss@ihep.ac.cn}
\thanks{Current address: Department of Physics and Institute of Theoretical
Physics, Nanjing Normal University, Nanjing 210023, China},
Shuang-Nan Zhang$^{2,4}$\thanks{E-mail: zhangsn@ihep.ac.cn}, Shu-Xu Yi$^{2}$, Yu Rong $^{2}$, Xu-Dong Gao $^{5,4}$ \\
$^1$\,Sabanc\i~University, Faculty of Engineering and Natural
Sciences, Orhanl\i$-$ Tuzla, \.{I}stanbul 34956, Turkey \\
$^2$\,Key Laboratory of Particle Astrophysics, Institute of High Energy
Physics, Chinese Academy of Sciences, Beijing 100049, China \\
$^3$\,Department of Physics, Xiangtan University, Xiangtan 411105, China \\
$^4$\,National Astronomical Observatories, Chinese Academy of Sciences,
Beijing, 100012, China \\
$^5$\,Department of Astronomy, Beijing Normal University, Beijing 100875, China \\
}

\date{}

\begin{document} \maketitle \label{firstpage}

\begin{abstract}
XTE J1810-189 underwent an outburst in 2008, and was observed over $\sim 100$ d
by {\it RXTE}. Performing a time-resolved spectral analysis on the photospheric
radius expansion burst detected on 2008 May 4, we obtain the source distance in
the range of 3.5--8.7 kpc for the first time. During its outburst, XTE
J1810-189 did not enter into the high/soft state, and both the soft and hard
colours decreased with decreasing flux. The fractional rms remained at high
values ($\sim 30$ per cent). The {\it RXTE}/PCA spectra for 3-25 keV can be
described by an absorbed power-law component with an additional Gaussian
component, and the derived photon index $\Gamma$ increased from $1.84\pm0.01$
to $2.25\pm0.04$ when the unabsorbed X-ray luminosity in 3-25 keV dropped from
$4\times10^{36}$ ergs s$^{-1}$ to $6\times10^{35}$ ergs s$^{-1}$. The
relatively high flux, dense observations and broadband spectra allow us to
provide strong evidence that the softening behaviour detected in the outburst
of XTE J1810-189 originates from the evolution of non-thermal component rather
than the thermal component (i.e. neutron star surface emission).

\end{abstract}


\section{Introduction}

X-ray binaries dominate the X-ray emission of our Galaxy, and hundreds of them
have been detected in the Galaxy and the Magellanic Clouds (Liu et al. 2006,
2007). A low-mass X-ray binary (XRB) is a system in which a low-mass companion
($M < M_{\odot}$) feeds a black-hole (BH) or a weakly magnetized neutron star
(NS) via Roche-lobe overflow. NS XRBs can be further classified into low
luminosity atoll sources ($\sim 0.001-0.5$ $L_{\rm Edd}$) and high luminosity Z
sources ($L_{\rm X} \sim L_{\rm Edd}$) according to their X-ray spectra and
timing properties (Hasinger \& van der Klis 1989; van der Klis 2006). Because
BHs and NSs have similar gravitational fields, it is difficult to distinguish
BH XRBs and NS XRBs unless type I X-ray bursts or coherent pulsations are
detected.

Type I X-ray bursts are sudden energy release processes caused by unstable
thermonuclear burning on the surface of NSs in XRBs (e.g. Lewin, van Paradijs
\& Taam 1993). The burst spectrum can be described well by a blackbody (BB)
component. In the special class burst events, the so-called photospheric radius
expansion (PRE) bursts, the peak X-ray luminosity reaches and remains at the
local Eddington luminosity for a few seconds, during which the atmosphere is
lifted up owing to strong radiation and the effective temperature decreases. At
the moment of touchdown, the photosphere settles on the NS surface again, i.e.
the radius derived from the BB normalization reaches a local minimum
corresponding to the radius of a NS. Since the mass and surface red shift of
NSs only vary in the narrow ranges, the Eddington luminosities measured from
PRE bursts are good distance indicators (Basinska et al. 1984; Kuulkers et al.
2003).

The colour-colour diagram (CCD), in which both soft and hard colours are
defined as the ratio of count rates between different energy bands, is
particularly useful for characterizing the behaviour of NS XRBs (e.g., Muno,
Remillard \& Chakrabarty 2002). The complete CCD from atoll sources has a Z
shape, consisting of, from top to bottom, the extreme island state, the island
state, and the banana state (e.g., Gierli{\'n}ski \& Done 2002). It has been
suggested that, as accretion rates increase, atoll sources go through from the
extreme island state to the banana state (e.g., Hasinger \& van der Klis 1989).
Following Lin, Remillard \& Homan (2007), we describe the extreme island,
island, banana states as low/hard, intermediate, and high/soft states, which
are consistent with BH XRBs. Some observational phenomena indicate that the
accretion rate might not monotonously increase when a source moves along the Z
pattern, especially in the branch for the intermediate state (Galloway et al.
2008). The fast time variability, which can be characterized by the root mean
square (rms), is another key tool for diagnosing accretion states in XRBs
(e.g., Mu\~{n}oz-Darias, Motta \& Belloni 2011; Heil, Vaughan \& Uttley 2012).
The positive correlation between rms and hardness (or colour) is commonly
observed in outbursts of both BH and NS XRBs (Lin et al. 2007; Mu\~{n}oz-Darias
et al. 2014). That is, XRBs have low variability when their radiation is
dominated by thermal emission and vise versa (Belloni 2010). However, this
correlation is not well explored below $\sim 0.01 L_{\rm Edd}$.

As already stated above, stellar-mass BHs and NSs have similar compactness;
therefore, BH XRBs and NS XRBs share many spectral and temporal properties at
high luminosity ($L_{\rm X} > 0.01 L_{\rm Edd}$; e.g., Done et al. 2007; Weng
\& Zhang 2011). Recently, much effort has been devoted to the differences
between two classes of XRB at low luminosity ($L_{\rm X} < 0.01 L_{\rm Edd}$):
(1) Compared to BH XRBs, NS XRBs are less radio loud at a given X-ray
luminosity, or in the other words, BH XRBs have a lower X-ray radiative
efficiency resulting from the advection-dominated accretion flows (Migliari \&
Fender 2006); (2) For the comparable accretion, NS XRBs are brighter than BH
XRBs by a factor of $\sim 100-1000$, owing to additional emission from their
solid surface (Narayan \& McClintock 2008); (3) As the luminosity decreases, it
has been proposed (Weng \& Zhang 2011) that accretion discs in NS XRBs begin to
leave the innermost stable circular orbit at higher luminosity, i.e. the discs
are truncated by the NS magnetospheres at low luminosity; (4) At
$10^{34}-10^{36}$ erg s$^{-1}$, NS XRBs have softer spectra, probably arising
from the soft thermal component emitted by NSs surface, which contributes a
significant fraction of X-ray emissions (Wijnands et al. 2014).

XTE J1810-189 underwent an outburst in 2008 and was monitored by {\it RXTE}
from 2008 March 19 to 2008 June 19. Markwardt, Strohmayer \& Swank (2008)
detected a type I X-ray burst from the source, identifying it as a NS XRB, and
also put an upper limit for distance of $\sim 11.5$ kpc according to the peak
flux of the type I X-ray burst. To date, the {\it RXTE} data have not been
systematically analyzed in the literature. In Section 2, we estimate the
interstellar hydrogen column density ($N_{\rm H}$) with {\it Swift}/XRT data,
measure the distance according to the mean peak flux of PRE burst detected by
{\it RXTE}, and also evaluate the spectral evolution of the 2008 outburst of
XTE J1810-189. We discuss these results in Section 3, and the summary follows
in Section 4.

\section{Data Analysis \& Results}

\subsection{Estimate absorption column density using {\it Swift} data}
Because {\it RXTE}/PCA is not sensitive to energies below 2 keV, the {\it
Swift}/XRT spectra are analysed to constrain the absorption column density.
Seven {\it Swift} pointing observations were made in 2008 March, and the other
one was made on 2011 June 19. The initial event cleaning is performed with the
task \texttt{xrtpipeline} with standard quality cuts. All XRT data were taken
in photon-counting mode; however, all of them suffer from problems of pile-up
because of high count rates $\sim$ 1.0-1.8 cts/s. We extract spectra within
annular regions centered on the source position with the inner radius $\sim$
6-8 pixels. The exposures maps are generated with \texttt{xrtexpomap} to
correct for bad columns, the ancillary response files are produced with the
task \texttt{xrtmkarf}, and the response matrix files (v014) are taken from the
CALDB database. To ensure valid results using $\chi^{2}$ statistical analysis,
the spectra are further grouped to have at least 30 counts per bin.

The {\it Swift}/XRT data in 2008 can be best fitted by an absorbed power-law
model with the photon index in the range of 1.4--1.9; however, the poor
statistics of the spectra do not permit two-component modelling. We fit these
spectra simultaneously and link the hydrogen column density to have a common
value, yielding $N_{\rm H} = 3.81^{+0.49}_{-0.46}\times10^{22}$ cm$^{-2}$
(Table \ref{xrt_fits}), consistent with the value reported by Starling, Kennea
\& Krimm (2008). The spectrum for 2011 June 19 is very soft with $\Gamma =
3.70_{-0.36}^{+0.40}$, indicating that the emissions are dominated by the
thermal component. Compared with the power-law model ($\chi^{2}/dof =
41.1/38$), the absorbed blackbody (BB) model provides much better fits to the
data ($\chi^{2}/dof = 31.5/38$), and the obtained $N_{\rm H} =
3.58^{+0.58}_{-0.50}\times10^{22}$ cm$^{-2}$ is consistent with the results
derived from the data in 2008. Thus, we adopt $N_{\rm H} =
3.81^{+0.49}_{-0.46}\times10^{22}$ cm$^{-2}$ for the following {\it RXTE}
spectral analysis.

\begin{table}\scriptsize
\centering \begin{tabular}{c c c c c c}
\hline Obs.ID & Obs-Date & $N_{\rm H}$ & $\Gamma/kT$ & $flux$ & $\chi^2$/dof \\
\hline
00031167001 &    2008-03-17& $3.81^{+0.49}_{-0.46}$  & $1.6_{-0.3}^{+0.3}$ & $4.0_{-0.5}^{+0.7}$ &169.8/169 \\
00306737000 &    2008-03-18&        ...              & $1.7_{-0.3}^{+0.3}$ & $4.9_{-0.6}^{+0.8}$ & ... \\
00031167002 &    2008-03-21&        ...              & $1.4_{-0.3}^{+0.3}$ & $5.0_{-0.5}^{+0.7}$ & ... \\
00031167003 &    2008-03-22&        ...              & $1.4_{-0.2}^{+0.2}$ & $5.6_{-0.5}^{+0.7}$ & ...  \\
00031167004 &    2008-03-23&        ...              & $1.4_{-0.2}^{+0.2}$ & $5.2_{-0.5}^{+0.6}$ & ...  \\
00031167005 &    2008-03-24&        ...              & $1.4_{-0.2}^{+0.2}$ & $6.6_{-0.6}^{+0.8}$ & ...  \\
00031167006 &    2008-03-25&        ...              & $1.9_{-0.3}^{+0.3}$ & $6.4_{-1.0}^{+1.4}$ & ...  \\
\hline
00455640000 &    2011-06-19& $3.58^{+0.58}_{-0.50}$  & $0.83_{-0.06}^{+0.06}$ & $2.7_{-0.3}^{+0.4}$ &31.5/38 \\
\hline
\end{tabular}
\caption{{\it Swift}/XRT spectra taken in 2008 are fitted by an absorbed
power-law model, and the spectrum in 2011 is fitted by an absorbed BB model
({\it phabs*bbodyrad} in XSPEC). $N_{\rm H}$: Absorption column density in
units of $10^{22}$ cm$^{-2}$. $flux$: 0.5--10 keV unabsorbed flux in units of
$10^{-10}$ erg cm$^{-2}$ s$^{-1}$. All errors are in 90\% confidence level.}
\label{xrt_fits}
\end{table}

\subsection{Properties of persistent emission}

\begin{figure}
\begin{center}
\includegraphics[width=9cm]{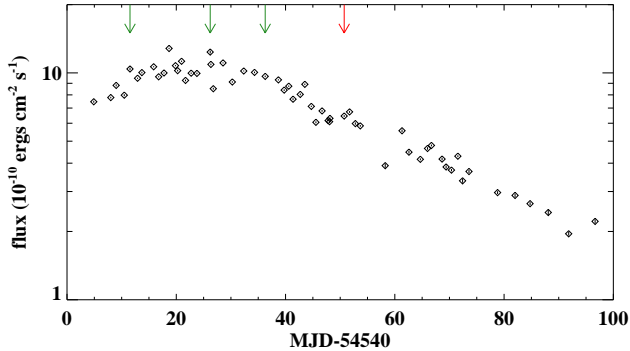}
\end{center}
\caption{The light curve of XTE J1810-189 during its 2008 outburst. The
unabsorbed flux in 3--25 keV is derived from the power-law fit (Table
\ref{po_fits}). The arrows label the observations in which the type I bursts
are detected, and the last (red) one corresponds to the PRE burst. \label{lc}}
\end{figure}

\begin{figure}
\begin{center}
\includegraphics[width=8cm]{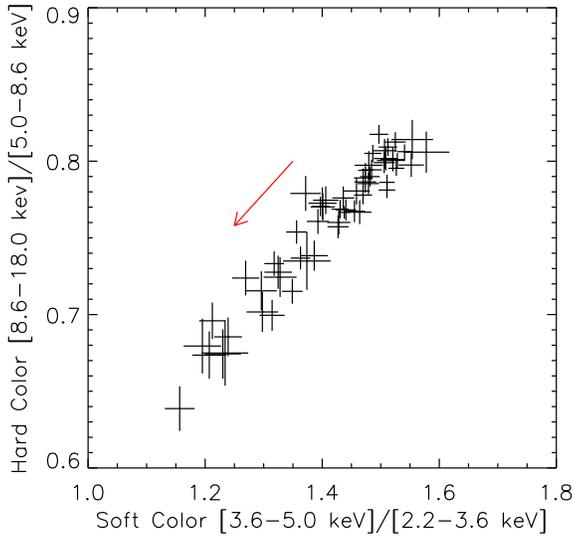}
\end{center}
\caption{Colour-colour diagram of XTE J1810-189. The red arrow labels the
direction of flux decay. \label{ccd}}
\end{figure}

\begin{figure}
\begin{center}
\includegraphics[width=8cm]{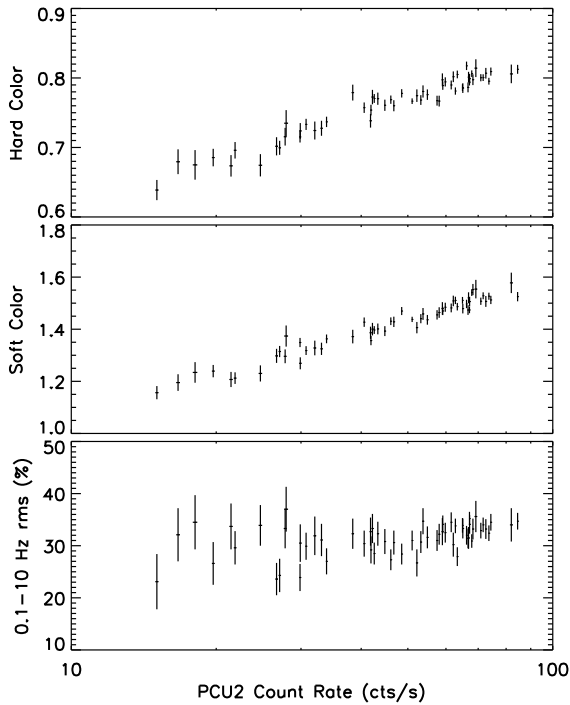}
\end{center}
\caption{Panels from top to bottom show the hard colour, soft colour, and
fractional rms plotted as the function of the PCU2 count rate (in 2.2--18.0
keV), respectively. \label{hid}}
\end{figure}

All 59 {\it RXTE} data of XTE J1810-189 taken by the Proportional Counter Array
(PCA) are analysed with the FTOOLS software package version 6.13. The
persistent emission, i.e. the X-ray emission when the type I X-ray bursts are
not present, is extracted from the Standard2 data from the top layer of PCU2
with the standard criteria: the Earth-limb elevation angle is larger than
$10\degr$ and the spacecraft pointing offset is less than $0.02\degr$. The
background files are created using the program \texttt{pcabackest}, while the
latest bright background model is adopted for the first 43 observations and the
faint background model is used for the last 16 observations, because the source
count rate is less than 40 counts s$^{-1}$ per PCU. Nevertheless, we checked
that our results are not sensitive to the background model. Type I X-ray bursts
are recorded in four observations (Fig. \ref{lc}), in which the data of 300 s
before and 1000 s after the type I X-ray bursts are excluded. The
background-subtracted light curves are created and their count rates are
averaged for each observation. The CCD of XTE J1810-189 is plotted in Fig.
\ref{ccd}. We define the soft colour as the ratio of the count rates in the PCA
channels of 9--11 (3.6--5.0 keV) and 0--8 (2.2--3.6 keV), and define the hard
colour as the ratio of the count rates in the PCA channels of 21--43 (8.6--18.0
keV) and 12--20 (5.0--8.6 keV), respectively. In the 2008 outburst, the source
evolved from the top right corner to the lower left corner, and both soft and
hard colours were tightly correlated with the intensity (in 2.2--18.0 keV, Fig.
\ref{hid}).

The light curves, in the same PCA channels (0--43) as those used for producing
the CCD, are extracted from the PCA Event mode data, E\_125us\_64M\_0\_1s for
timing analysis. The power density spectra (PDS) are constructed using light
curve segments of 32 s and 8 ms time bins with the task \texttt{powspec}
(version 1.0). We adopt the method of Miyamoto to normalize the PDSs (Miyamoto,
Kimura \& Kitamoto 1991), and average them with the logarithmic rebin option
-1.03 for each observation. The Poisson noise is subtracted (\texttt{norm =
-2}), and the fractional rms is integrated over the 0.1--10 Hz frequency band.
As can be seen in Fig. \ref{hid}, XTE J1810-189 displays a high and nearly
constant variability level ($\sim 30$ per cent rms) during the outburst.

\begin{table} \scriptsize
\centering \begin{tabular}{c c c c c c}
\hline Obs.ID & MJD & Exp.(s) & $\Gamma$ & Flux & $\chi^2$/dof \\
\hline
93044-07-03-00 &        54544.9& 6080 & $1.94_{-0.02}^{+0.02}$ & $7.46_{-0.07}^{+0.07}$ &100.5/45 \\
93044-07-04-00 &        54548.0& 1968 & $1.95_{-0.02}^{+0.02}$ & $7.79_{-0.06}^{+0.06}$ &68.9/45 \\
93044-07-04-01 &        54549.0&  992 & $1.89_{-0.02}^{+0.02}$ & $8.81_{-0.09}^{+0.09}$ &53.1/45 \\
93044-07-04-02 &        54550.5& 1248 & $1.90_{-0.02}^{+0.02}$ & $7.98_{-0.08}^{+0.08}$ &30.6/45 \\
93044-07-05-00 &        54552.9& 2624 & $1.88_{-0.01}^{+0.01}$ & $9.47_{-0.06}^{+0.06}$ &48.2/45 \\
93044-07-05-01 &        54551.5&  448 & $1.87_{-0.03}^{+0.03}$ & $10.40_{-0.14}^{+0.14}$ &41.3/45 \\
93044-07-06-00 &        54553.7& 1232 & $1.87_{-0.02}^{+0.02}$ & $10.03_{-0.09}^{+0.09}$ &42.5/45 \\
93433-01-01-00 &        54555.9& 2608 & $1.86_{-0.01}^{+0.01}$ & $10.64_{-0.06}^{+0.06}$ &62.8/45 \\
93433-01-01-01 &        54556.8& 3184 & $1.92_{-0.01}^{+0.01}$ & $9.62_{-0.06}^{+0.06}$ &85.1/45 \\
93433-01-01-02 &        54557.8& 3200 & $1.90_{-0.01}^{+0.01}$ & $10.00_{-0.06}^{+0.06}$ &66.2/45 \\
93433-01-01-03 &        54558.7& 1376 & $1.84_{-0.01}^{+0.01}$ & $12.81_{-0.09}^{+0.09}$ &31.6/45 \\
93433-01-01-04 &        54559.8& 2912 & $1.87_{-0.01}^{+0.01}$ & $10.76_{-0.06}^{+0.06}$ &62.2/45 \\
93433-01-02-00 &        54560.3& 1232 & $1.89_{-0.02}^{+0.02}$ & $10.23_{-0.09}^{+0.09}$ &72.3/45 \\
93433-01-02-01 &        54561.0& 1920 & $1.85_{-0.01}^{+0.01}$ & $11.25_{-0.07}^{+0.07}$ &54.4/45 \\
93433-01-02-02 &        54561.7& 1408 & $1.89_{-0.02}^{+0.02}$ & $9.27_{-0.08}^{+0.08}$ &47.7/45 \\
93433-01-02-03 &        54562.7& 2064 & $1.86_{-0.01}^{+0.01}$ & $9.98_{-0.07}^{+0.07}$ &55.3/45 \\
93433-01-02-04 &        54563.8& 1440 & $1.88_{-0.02}^{+0.02}$ & $9.95_{-0.08}^{+0.08}$ &50.8/45 \\
93433-01-02-05 &        54566.2&  320 & $1.84_{-0.03}^{+0.03}$ & $12.35_{-0.18}^{+0.18}$ &50.0/45 \\
93433-01-02-06 &        54566.3& 1136 & $1.86_{-0.02}^{+0.02}$ & $10.90_{-0.09}^{+0.09}$ &52.3/45 \\
93433-01-02-07 &        54566.8& 1376 & $1.95_{-0.02}^{+0.02}$ & $8.52_{-0.08}^{+0.08}$ &61.9/45 \\
93433-01-03-00 &        54568.6& 2720 & $1.87_{-0.01}^{+0.01}$ & $11.07_{-0.06}^{+0.06}$ &73.7/45 \\
93433-01-03-01 &        54570.3& 1952 & $1.91_{-0.01}^{+0.01}$ & $9.12_{-0.07}^{+0.07}$ &49.4/45 \\
93433-01-03-02 &        54572.3& 3312 & $1.87_{-0.01}^{+0.01}$ & $10.19_{-0.06}^{+0.06}$ &69.5/45 \\
93433-01-04-00 &        54574.3& 1552 & $1.88_{-0.01}^{+0.01}$ & $10.05_{-0.08}^{+0.08}$ &49.2/45 \\
93433-01-04-01 &        54576.3& 1552 & $1.91_{-0.01}^{+0.02}$ & $9.66_{-0.08}^{+0.08}$ &47.5/45 \\
93433-01-04-02 &        54578.7& 2896 & $1.91_{-0.01}^{+0.01}$ & $9.32_{-0.06}^{+0.06}$ &88.7/45 \\
93433-01-04-03 &        54579.8& 1728 & $1.97_{-0.02}^{+0.02}$ & $8.40_{-0.07}^{+0.07}$ &74.3/45 \\
93433-01-04-04 &        54580.6& 2288 & $1.92_{-0.01}^{+0.01}$ & $8.74_{-0.06}^{+0.06}$ &56.2/45 \\
93433-01-05-00 &        54581.4& 1248 & $1.95_{-0.02}^{+0.02}$ & $7.66_{-0.08}^{+0.08}$ &47.1/45 \\
93433-01-05-01 &        54582.7& 1664 & $1.95_{-0.02}^{+0.02}$ & $8.04_{-0.07}^{+0.07}$ &64.2/45 \\
93433-01-05-02 &        54583.5& 1952 & $1.89_{-0.01}^{+0.01}$ & $8.91_{-0.07}^{+0.07}$ &60.9/45 \\
93433-01-05-03 &        54584.7& 3040 & $1.94_{-0.01}^{+0.01}$ & $7.12_{-0.05}^{+0.05}$ &68.0/45 \\
93433-01-05-04 &        54585.6& 2208 & $1.98_{-0.02}^{+0.02}$ & $6.06_{-0.06}^{+0.06}$ &101.5/45 \\
93433-01-05-05 &        54586.7& 1760 & $1.96_{-0.02}^{+0.02}$ & $6.80_{-0.06}^{+0.06}$ &56.8/45 \\
93433-01-05-06 &        54587.8& 3200 & $1.98_{-0.01}^{+0.01}$ & $6.16_{-0.05}^{+0.05}$ &77.0/45 \\
93433-01-06-00 &        54588.1& 1280 & $1.98_{-0.02}^{+0.02}$ & $6.10_{-0.07}^{+0.07}$ &64.4/45 \\
93433-01-06-01 &        54588.1& 1680 & $1.97_{-0.02}^{+0.02}$ & $6.31_{-0.06}^{+0.06}$ &58.5/45 \\
93433-01-06-02 &        54590.7& 1824 & $1.99_{-0.02}^{+0.02}$ & $6.46_{-0.06}^{+0.06}$ &45.8/45 \\
93433-01-06-03 &        54591.7& 3200 & $1.95_{-0.02}^{+0.02}$ & $6.73_{-0.07}^{+0.07}$ &100.5/45 \\
93433-01-06-04 &        54592.8& 1312 & $2.00_{-0.02}^{+0.02}$ & $5.98_{-0.07}^{+0.07}$ &48.5/45 \\
93433-01-06-05 &        54593.7& 2416 & $2.00_{-0.02}^{+0.02}$ & $5.84_{-0.05}^{+0.05}$ &74.6/45 \\
93433-01-07-00 &        54598.2&  656 & $2.09_{-0.04}^{+0.04}$ & $3.90_{-0.08}^{+0.08}$ &49.8/45 \\
93433-01-07-02 &        54601.3& 1184 & $1.96_{-0.02}^{+0.02}$ & $5.56_{-0.07}^{+0.07}$ &42.8/45 \\
93433-01-08-00 &        54602.6& 1104 & $2.08_{-0.03}^{+0.03}$ & $4.47_{-0.07}^{+0.07}$ &34.3/45 \\
93433-01-08-01 &        54604.7& 3184 & $2.10_{-0.02}^{+0.02}$ & $4.16_{-0.04}^{+0.04}$ &55.3/45 \\
93433-01-08-02 &        54606.7& 3168 & $2.05_{-0.02}^{+0.02}$ & $4.79_{-0.04}^{+0.04}$ &62.3/45 \\
93433-01-08-03 &        54608.7& 1664 & $2.07_{-0.03}^{+0.03}$ & $4.17_{-0.06}^{+0.06}$ &59.8/45 \\
93433-01-08-04 &        54606.0& 1536 & $2.07_{-0.02}^{+0.02}$ & $4.65_{-0.06}^{+0.06}$ &47.3/45 \\
93433-01-09-00 &        54609.4& 1392 & $2.11_{-0.03}^{+0.03}$ & $3.85_{-0.06}^{+0.06}$ &40.0/45 \\
93433-01-09-01 &        54610.4& 2224 & $2.14_{-0.02}^{+0.02}$ & $3.73_{-0.05}^{+0.05}$ &47.3/45 \\
93433-01-09-02 &        54611.5& 3200 & $2.08_{-0.02}^{+0.02}$ & $4.29_{-0.04}^{+0.04}$ &61.5/45 \\
93433-01-09-03 &        54612.4& 1008 & $2.19_{-0.03}^{+0.04}$ & $3.34_{-0.06}^{+0.06}$ &35.9/45 \\
93433-01-09-04 &        54613.6& 1408 & $2.13_{-0.03}^{+0.03}$ & $3.68_{-0.06}^{+0.06}$ &48.8/45 \\
93433-01-10-00 &        54618.8& 2288 & $2.14_{-0.03}^{+0.03}$ & $2.97_{-0.04}^{+0.04}$ &51.2/45 \\
93433-01-10-02 &        54622.0& 1296 & $2.18_{-0.04}^{+0.04}$ & $2.88_{-0.05}^{+0.05}$ &41.2/45 \\
93433-01-11-00 &        54624.8& 2400 & $2.19_{-0.03}^{+0.03}$ & $2.65_{-0.04}^{+0.04}$ &49.9/45 \\
93433-01-11-01 &        54628.1&  960 & $2.21_{-0.05}^{+0.05}$ & $2.42_{-0.06}^{+0.06}$ &29.3/45 \\
93433-01-12-00 &        54631.8& 2432 & $2.25_{-0.04}^{+0.04}$ & $1.95_{-0.04}^{+0.04}$ &49.3/45 \\
93433-01-12-01 &        54636.7& 1488 & $2.21_{-0.04}^{+0.04}$ & $2.22_{-0.05}^{+0.05}$ &45.2/45 \\
\hline
\end{tabular}
\caption{Spectra are fitted by an absorbed power-law model. Exp: Exposure time
in units of second. Flux: 3--25 keV unabsorbed flux in units of $10^{-10}$ erg
cm$^{-2}$ s$^{-1}$. All errors are for the 90\% confidence level.}
\label{po_fits}
\end{table}

\begin{figure}
\includegraphics[width=9cm]{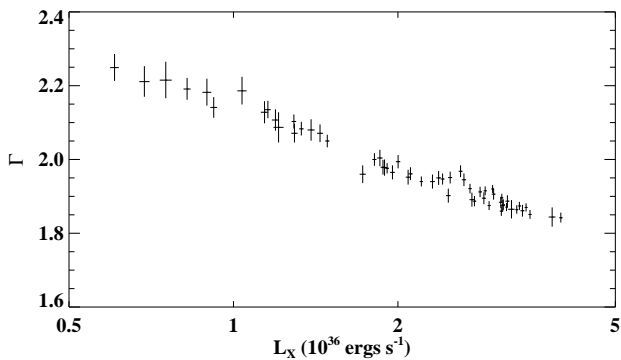}
\caption{Spectra soften when the X-ray luminosity ($d =5.1$ kpc is used)
decrease. \label{gamma}}
\end{figure}

The response matrix files are created by the generator PCARMF (v11.7), and the
recommended systematic error of $0.5\%$ is applied for the spectral analysis.
The persistent emission is fitted for 3-25 keV by an absorbed power-law model
with the $N_{\rm H}$ fixed to $3.81\times10^{22}$ cm$^{-2}$. An additional
Gaussian component is required to mimic an iron line at $\sim 6.4$ keV. This
model adequately fits all spectra with the reduced $\chi^{2}$ values in the
range $0.65-2.26$. Only three observations with relatively long exposure time
(observation IDs: 93044-7-03-00, 93433-01-05-04, and 93433-01-06-03) have the
reduced $\chi^{2}$ larger than 2. The worse fits might be due to strong
spectral fluctuation; therefore, we assign a typical relative error of $\sim
1\%$ to $\Gamma$ for these data (Table \ref{po_fits}). The unabsorbed flux is
calculated with the convolution model, {\it cflux}. As can be seen in Fig.
\ref{gamma}, the spectra become softer when the X-ray flux decreases.

\subsection{Evaluate source distance from PRE burst}

The PCA Event mode data are analysed to study the PRE burst detected on 2008
May 4 (observation ID = 93433-01-06-02). We extract the time-resolved spectra
of PRE burst within a time bin size of 0.25 s around the burst peak, and use
the longer integration time to compensate for the flux decay in the burst tail.
Since the count rate of persistent emission is relatively low ($\sim 60$
cts/s), we extract the spectrum from the whole observation in burst-free
intervals to minimize the fluctuation, and the spectrum is used as the
background for the PRE burst spectral modelling. The dead time effect is
corrected following the approach suggested by the {\it RXTE} team
\footnote{http://heasarc.nasa.gov/docs/xte/recipes/pca\_deadtime.html}. The
spectra are fitted by a BB model ({\it phabs*bbodyrad} in XSPEC), and the C
statistic is used to accommodate low count rates. The fittings are generally
performed over the energy range of 3-20 keV but concentrated to 3-18 keV or
3-15 keV to avoid unphysical parameters (mostly for the spectra at the late
stage). For this observation, only PCU2 was on and channel 11 had zero counts;
therefore, this channel is ignored in the fitting.

\begin{figure}
\includegraphics[width=9cm]{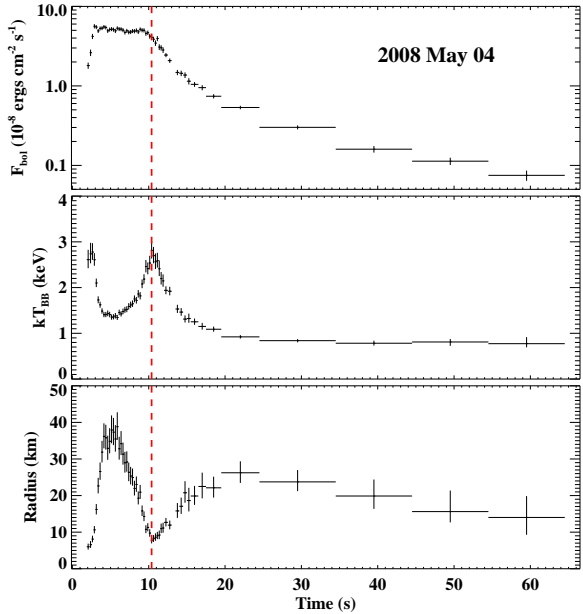}
\caption{Time-resolved spectra of PRE burst are fitted by the absorbed BB
model, and the radius of BB emission is derived by assuming $d = 10$ kpc. The
red dashed lines mark the onset of touchdown. \label{pre}}
\end{figure}

The derived parameters are plotted in Fig. \ref{pre}, in which the red dashed
lines mark the moment of touchdown. The bolometric flux of a burst is evaluated
from the {\it bbodyrad} model as:
\begin{equation}
F_{\rm bol} = 1.076\times10^{-11}\ kT_{\rm BB}^{4} \ N_{\rm BB} \quad  {\rm
ergs \ cm^{-2} \ s^{-1}},
\end{equation}
where $kT_{\rm BB}$ is the BB temperature in units of keV, and $N_{\rm BB}$ is
the normalization. The measured flux is lower than the locally observed value
by a factor of $(1+z(R))^{-2}$ due to the gravitational red shift, where the
red-shift factor $z(R) = \frac{1}{\sqrt{1-\frac{2GM}{Rc^{2}}}}-1$ (Lewin et al.
1993). Because the photosphere is lifted up to higher radius (smaller $z(R)$),
the mean peak flux of PRE burst $F_{\rm peak,PRE} = (5.09\pm0.30)\times10^{-8}$
ergs cm$^{-2}$ s$^{-1}$ is slightly larger than the touchdown flux: $F_{\rm TD}
= (4.08\pm0.31)\times10^{-8}$ ergs cm$^{-2}$ s$^{-1}$. Since there is only a
small discrepancy between $F_{\rm peak,PRE}$ and $F_{\rm TD}$, we adopt the
standard process, i.e. using $F_{\rm peak,PRE}$ as the Eddington luminosity
(from the NS surface) to minimize the fluctuation of spectral fitting.

We calculate the distance of XTE J1810-189 according to
\begin{equation}
\begin{aligned}
d = 8.6(\frac{F_{\rm peak,PRE}}{3\times10^{-8} \ {\rm ergs \ cm^{-2} \
s^{-1}}})^{-1/2} (\frac{M_{\rm NS}}{1.4 \ M_{\odot}})^{1/2} \\
\times[\frac{1+z(R)}{1.31}]^{-1/2}(1+X)^{-1/2} \quad {\rm kpc},
\end{aligned}
\end{equation}
given by Galloway et al. (2008), where $M_{\rm NS}$ is the mass of NS and $X$
is the hydrogen mass fraction in the atmosphere. Assuming the typical values
$M_{\rm NS} = 1.4 M_{\odot}$, $z(R) =0.31$ (for $M_{\rm NS} = 1.4 M_{\odot}$
and $R_{\rm NS} = 10$ km), and $X = 0.7$ (cosmic abundances), we obtain $d =
5.1\pm0.2$ kpc, that is significantly smaller than the upper limit ($\sim 11.5$
kpc) reported by Markwardt et al. (2008).

\section{Discussions}

Fig. \ref{hid} shows that both the soft and hard colours decrease as the flux
decays, and the fractional rms remains at $\sim 30$ per cent. Even though the
values of both colours are consistent with an intermediate state, both the low
luminosity and the strong variability point to a low/hard state for the atoll
source. Accretion states of XRBs have been well explored at high luminosity,
and colour variations are interpreted as the coevolution of both thermal and
non-thermal components. Generally, compared to cases in the high/soft state,
the accretion disc in the low/hard state is cooler, the power-law component
becomes harder (smaller $\Gamma$), and the power-law fraction increases (e.g.
Dunn et al. 2010; Mu\~{n}oz-Darias et al. 2013). However, a study of the
spectral evolution at low luminosity, when the contribution of accretion disc
is negligible, is challenged by the low flux. Currently, collecting a sample of
spectra at 0.5--10 keV for NS XRBs (i.e. taken from {\it Chandra}/{\it
XMM-Newton}/{\it Swift}), Wijnands et al. (2014) suggested that the photon
index increases with decreasing X-ray luminosity between $10^{34}$ and
$10^{36}$ ergs s$^{-1}$, and BH XRBs have significantly harder spectra at the
same luminosity (Armas Padilla et al. 2011; Plotkin, Gallo \& Jonker 2013).
However, because of the sparse data points for each source and large scatters
on $\Gamma$, the anti-correlation between $\Gamma$ and the X-ray luminosity is
relatively weak. Moreover, it is difficult to reveal the nature of spectral
softening with the limited band spectra (0.5--10 keV), and two possible
scenarios have been proposed (see also Allen et al. 2015): (1) The thermal
emission from the NS surface becomes the major component at the low luminosity.
(2) The power-law component softens as the X-ray luminosity decrease to
$10^{35}$ erg s$^{-1}$, below which the thermal component and another harder
power-law component start to overcome the original power-law emission.

In this work, we reveal that the significant softening process occurs in the
2008 outburst of XTE J1810-189 in the flux range of $(2-13) \times 10^{-10}$
ergs cm$^{-2}$ s$^{-1}$, which is well above the PCA background level. Since
the PCA spectra in 3--25 keV provide a good constraint on the power-law
component, the tight correlation between $\Gamma$ and the X-ray luminosity with
the dense {\it RXTE} observations is found in Fig. \ref{gamma}. Fitting all
these data with the function $\Gamma = a\ {\rm log} L_{\rm X}+b$ using the
error of $\Gamma$ as the weight, we obtain the coefficients: $a =
-0.52\pm0.01$, $b = 20.8\pm0.5$. The source has $\Gamma \geq 2.1$ at $L_{\rm X}
\leq 10^{36}$ ergs s$^{-1}$, that is softer than spectra of most BH XRBs at the
same luminosity.

To check whether the softening spectra result from a larger contribution of NS
surface emissions, we refit the data using the model with an additional BB
component included ({\it phabs*(powerlaw+bbodyrad+gau)} in XSPEC). Because {\it
RXTE}/PCA is not sensitive at the energy band below 3 keV, the surface
temperature $kT$ is allowed to vary, but restricted to below 5 keV to avoid
unphysical parameters (Table \ref{bb_fits}). As can be seen, the thermal
component only contributes a small fraction ($< 10\%$) of the total flux and
becomes cooler as the flux decays. Since the main part of BB component with $kT
\leq 1$ keV is out of the coverage of PCA, its unabsorbed flux in 3--25 keV is
less constrained. When the thermal component is added, the non-thermal
component becomes slightly harder with a larger error in $\Gamma$. Nonetheless,
the softening trend still emerges in the fitting; that is, $\Gamma$ increases
from $1.67_{-0.07}^{+0.06}$ to $2.10_{-0.15}^{+0.09}$. The high and constant
variability levels indicate that the emissions in 3--25 keV are dominated by
the non-thermal component, and the fractional contribution of the thermal
component does not increase with decreasing flux. Note that the model's
independent parameter, the hard colour ((8.6--18.0)/(5.0--8.6) keV) also
decreases following the flux decays (Fig. \ref{hid}). Therefore, we demonstrate
that the softer spectra at lower luminosity are caused by the softening of the
power-law component.

The largest uncertainty in our results arises from the distance of the source,
which depends on $M_{\rm NS}$, $R_{\rm NS}$ ($z$), and the composition of
accretion material $X$. The caveats for using PRE bursts as the distance
indicator are: (1) The Eddington luminosities vary from source to source
(Kuulkers et al. 2003) and (2) even in the same source, PRE bursts could reach
different peak fluxes due to the variation of burst fuel composition (Galloway
et al. 2008). Assuming that $M_{\rm NS}$, $R_{\rm NS}$, and $X$ are uniformly
distributed in the typical range of $0.8-2.0 M_{\odot}$, $8-16$ km (e.g.
Lattimer \& Prakash. 2007; Steiner, Lattimer \& Brown 2010), and $0-1$,
respectively, while $F_{\rm peak,PRE} = (5.09\pm0.30)\times10^{-8}$ ergs
cm$^{-2}$ having a normal distribution, we randomly generate 100 points for
each parameters, then calculate $10^{8}$ of $d$ according to Equation (2). The
simulated $d$ has values in the range of $3.5-8.7$ kpc, and the Gaussian fit to
the profile of its distribution indicates $d = (5.5\pm1.7)$ kpc. Thus, we
suggest that the $L_{\rm X}$ used above could vary by a factor of 3 at most,
which would not change our conclusions.

\begin{table*} \scriptsize
\centering \begin{tabular}{c c c c c c c}
\hline Obs.ID & MJD & $\Gamma$ & $kT$ (keV) & $f_{\rm BB}$& Flux & $\chi^2$/dof \\
\hline
93044-07-03-00 &        54544.9&  $1.77_{-0.04}^{+0.04}$ & $1.01_{-0.07}^{+0.06}$ & $0.56_{-0.13}^{+0.14}$ & $7.63_{-0.04}^{+0.04}$ &38.1/43 \\
93044-07-04-00 &        54548.0&  $1.75_{-0.06}^{+0.06}$ & $0.94_{-0.09}^{+0.08}$ & $0.60_{-0.18}^{+0.19}$ & $7.99_{-0.07}^{+0.07}$ &33.7/43 \\
93044-07-04-01 &        54549.0&  $1.79_{-0.08}^{+0.08}$ & $1.00_{-0.33}^{+0.19}$ & $0.40_{-0.29}^{+0.32}$ & $8.95_{-0.10}^{+0.10}$ &47.6/43 \\
93044-07-04-02 &        54550.5&  $1.77_{-0.08}^{+0.08}$ & $1.14_{-0.15}^{+0.13}$ & $0.51_{-0.30}^{+0.30}$ & $8.09_{-0.09}^{+0.09}$ &22.7/43 \\
93044-07-05-00 &        54552.9&  $1.76_{-0.05}^{+0.05}$ & $1.02_{-0.13}^{+0.10}$ & $0.50_{-0.20}^{+0.21}$ & $9.56_{-0.06}^{+0.06}$ &28.0/43 \\
93044-07-05-01 &        54551.5&  $1.70_{-0.11}^{+0.10}$ & $1.06_{-0.20}^{+0.15}$ & $0.72_{-0.44}^{+0.47}$ & $10.60_{-0.16}^{+0.16}$ &33.6/43 \\
93044-07-06-00 &        54553.7&  $1.74_{-0.07}^{+0.07}$ & $1.11_{-0.13}^{+0.11}$ & $0.63_{-0.31}^{+0.31}$ & $10.18_{-0.09}^{+0.09}$ &30.4/43 \\
93433-01-01-00 &        54555.9&  $1.72_{-0.05}^{+0.05}$ & $1.10_{-0.09}^{+0.07}$ & $0.70_{-0.23}^{+0.24}$ & $10.82_{-0.07}^{+0.07}$ &34.7/43 \\
93433-01-01-01 &        54556.8&  $1.75_{-0.05}^{+0.04}$ & $1.07_{-0.07}^{+0.06}$ & $0.73_{-0.19}^{+0.20}$ & $9.78_{-0.06}^{+0.06}$ &42.5/43 \\
93433-01-01-02 &        54557.8&  $1.77_{-0.04}^{+0.04}$ & $1.02_{-0.11}^{+0.09}$ & $0.54_{-0.19}^{+0.20}$ & $10.16_{-0.06}^{+0.06}$ &38.9/43 \\
93433-01-01-03 &        54558.7&  $1.75_{-0.06}^{+0.06}$ & $1.18_{-0.16}^{+0.14}$ & $0.58_{-0.35}^{+0.35}$ & $12.96_{-0.10}^{+0.10}$ &23.9/43 \\
93433-01-01-04 &        54559.8&  $1.74_{-0.05}^{+0.04}$ & $1.12_{-0.09}^{+0.08}$ & $0.67_{-0.23}^{+0.23}$ & $10.92_{-0.06}^{+0.07}$ &36.0/43 \\
93433-01-02-00 &        54560.3&  $1.67_{-0.07}^{+0.06}$ & $1.01_{-0.09}^{+0.08}$ & $0.91_{-0.27}^{+0.28}$ & $10.49_{-0.10}^{+0.10}$ &34.7/43 \\
93433-01-02-01 &        54561.0&  $1.70_{-0.05}^{+0.05}$ & $1.14_{-0.09}^{+0.08}$ & $0.80_{-0.27}^{+0.27}$ & $11.41_{-0.08}^{+0.08}$ &28.7/43 \\
93433-01-02-02 &        54561.7&  $1.72_{-0.07}^{+0.07}$ & $1.06_{-0.11}^{+0.09}$ & $0.70_{-0.27}^{+0.28}$ & $9.43_{-0.09}^{+0.09}$ &27.1/43 \\
93433-01-02-03 &        54562.7&  $1.76_{-0.05}^{+0.05}$ & $1.00_{-0.16}^{+0.12}$ & $0.44_{-0.21}^{+0.23}$ & $10.13_{-0.07}^{+0.07}$ &40.6/43 \\
93433-01-02-04 &        54563.8&  $1.73_{-0.07}^{+0.07}$ & $1.11_{-0.12}^{+0.10}$ & $0.70_{-0.30}^{+0.30}$ & $10.11_{-0.09}^{+0.09}$ &34.9/43 \\
93433-01-02-05 &        54566.2&  $1.73_{-0.12}^{+0.10}$ & $1.31_{-0.22}^{+0.34}$ & $0.86_{-0.65}^{+0.67}$ & $12.51_{-0.19}^{+0.19}$ &45.3/43 \\
93433-01-02-06 &        54566.3&  $1.70_{-0.06}^{+0.06}$ & $1.03_{-0.12}^{+0.10}$ & $0.74_{-0.29}^{+0.30}$ & $11.13_{-0.10}^{+0.10}$ &30.8/43 \\
93433-01-02-07 &        54566.8&  $1.76_{-0.07}^{+0.07}$ & $1.05_{-0.11}^{+0.09}$ & $0.69_{-0.27}^{+0.27}$ & $8.60_{-0.08}^{+0.08}$ &41.8/43 \\
93433-01-03-00 &        54568.6&  $1.69_{-0.05}^{+0.05}$ & $1.13_{-0.06}^{+0.06}$ & $0.95_{-0.23}^{+0.23}$ & $11.26_{-0.07}^{+0.07}$ &25.2/43 \\
93433-01-03-01 &        54570.3&  $1.80_{-0.06}^{+0.05}$ & $1.02_{-0.14}^{+0.11}$ & $0.47_{-0.22}^{+0.24}$ & $9.27_{-0.07}^{+0.07}$ &35.5/43 \\
93433-01-03-02 &        54572.3&  $1.74_{-0.04}^{+0.04}$ & $1.04_{-0.09}^{+0.08}$ & $0.58_{-0.19}^{+0.19}$ & $10.37_{-0.06}^{+0.06}$ &38.1/43 \\
93433-01-04-00 &        54574.3&  $1.72_{-0.06}^{+0.06}$ & $1.06_{-0.11}^{+0.09}$ & $0.68_{-0.26}^{+0.27}$ & $10.24_{-0.08}^{+0.08}$ &27.8/43 \\
93433-01-04-01 &        54576.3&  $1.79_{-0.06}^{+0.06}$ & $1.02_{-0.17}^{+0.12}$ & $0.49_{-0.26}^{+0.27}$ & $9.82_{-0.08}^{+0.08}$ &35.4/43 \\
93433-01-04-02 &        54578.7&  $1.70_{-0.05}^{+0.05}$ & $1.01_{-0.07}^{+0.06}$ & $0.82_{-0.18}^{+0.19}$ & $9.54_{-0.06}^{+0.06}$ &20.6/43 \\
93433-01-04-03 &        54579.8&  $1.75_{-0.06}^{+0.06}$ & $0.96_{-0.09}^{+0.07}$ & $0.75_{-0.21}^{+0.22}$ & $8.61_{-0.08}^{+0.08}$ &30.8/43 \\
93433-01-04-04 &        54580.6&  $1.80_{-0.05}^{+0.05}$ & $0.94_{-0.14}^{+0.11}$ & $0.42_{-0.18}^{+0.20}$ & $8.90_{-0.07}^{+0.07}$ &37.8/43 \\
93433-01-05-00 &        54581.4&  $1.80_{-0.08}^{+0.08}$ & $0.97_{-0.15}^{+0.12}$ & $0.48_{-0.24}^{+0.26}$ & $7.80_{-0.09}^{+0.09}$ &34.8/43 \\
93433-01-05-01 &        54582.7&  $1.78_{-0.06}^{+0.06}$ & $0.93_{-0.12}^{+0.10}$ & $0.54_{-0.20}^{+0.21}$ & $8.22_{-0.08}^{+0.08}$ &40.1/43 \\
93433-01-05-02 &        54583.5&  $1.71_{-0.06}^{+0.06}$ & $1.00_{-0.10}^{+0.08}$ & $0.65_{-0.20}^{+0.21}$ & $9.11_{-0.07}^{+0.07}$ &27.0/43 \\
93433-01-05-03 &        54584.7&  $1.80_{-0.06}^{+0.05}$ & $0.97_{-0.12}^{+0.10}$ & $0.42_{-0.16}^{+0.17}$ & $7.28_{-0.06}^{+0.06}$ &45.0/43 \\
93433-01-05-04 &        54585.6&  $1.78_{-0.07}^{+0.07}$ & $0.83_{-0.13}^{+0.10}$ & $0.44_{-0.15}^{+0.16}$ & $6.26_{-0.06}^{+0.06}$ &62.3/43 \\
93433-01-05-05 &        54586.7&  $1.79_{-0.07}^{+0.07}$ & $0.95_{-0.14}^{+0.11}$ & $0.46_{-0.20}^{+0.21}$ & $6.98_{-0.07}^{+0.07}$ &37.7/43 \\
93433-01-05-06 &        54587.8&  $1.79_{-0.05}^{+0.05}$ & $0.84_{-0.09}^{+0.08}$ & $0.43_{-0.11}^{+0.12}$ & $6.34_{-0.05}^{+0.05}$ &24.3/43 \\
93433-01-06-00 &        54588.1&  $1.77_{-0.08}^{+0.08}$ & $0.77_{-0.13}^{+0.12}$ & $0.44_{-0.16}^{+0.18}$ & $6.32_{-0.08}^{+0.08}$ &32.5/43 \\
93433-01-06-01 &        54588.1&  $1.82_{-0.07}^{+0.07}$ & $0.80_{-0.18}^{+0.14}$ & $0.32_{-0.15}^{+0.18}$ & $6.47_{-0.07}^{+0.07}$ &40.0/43 \\
93433-01-06-02 &        54590.7&  $1.82_{-0.07}^{+0.07}$ & $0.85_{-0.13}^{+0.11}$ & $0.42_{-0.16}^{+0.18}$ & $6.63_{-0.07}^{+0.07}$ &20.1/43 \\
93433-01-06-03 &        54591.7&  $1.71_{-0.06}^{+0.05}$ & $0.98_{-0.07}^{+0.06}$ & $0.64_{-0.15}^{+0.15}$ & $6.92_{-0.05}^{+0.05}$ &37.1/43 \\
93433-01-06-04 &        54592.8&  $1.73_{-0.09}^{+0.09}$ & $0.98_{-0.09}^{+0.08}$ & $0.67_{-0.21}^{+0.22}$ & $6.18_{-0.08}^{+0.08}$ &20.2/43 \\
93433-01-06-05 &        54593.7&  $1.79_{-0.06}^{+0.06}$ & $0.87_{-0.09}^{+0.08}$ & $0.47_{-0.14}^{+0.15}$ & $6.03_{-0.06}^{+0.06}$ &33.2/43 \\
93433-01-07-00 &        54598.2&  $1.80_{-0.16}^{+0.14}$ & $0.77_{-0.16}^{+0.14}$ & $0.39_{-0.19}^{+0.21}$ & $4.08_{-0.10}^{+0.10}$ &35.1/43 \\
93433-01-07-02 &        54601.3&  $1.80_{-0.08}^{+0.08}$ & $0.72_{-0.17}^{+0.14}$ & $0.30_{-0.14}^{+0.16}$ & $5.75_{-0.08}^{+0.08}$ &23.3/43 \\
93433-01-08-00 &        54602.6&  $1.92_{-0.11}^{+0.10}$ & $0.71_{-0.24}^{+0.17}$ & $0.25_{-0.15}^{+0.17}$ & $4.63_{-0.08}^{+0.08}$ &22.7/43 \\
93433-01-08-01 &        54604.7&  $1.99_{-0.06}^{+0.06}$ & $0.65_{-0.16}^{+0.13}$ & $0.16_{-0.08}^{+0.10}$ & $4.28_{-0.05}^{+0.05}$ &38.0/43 \\
93433-01-08-02 &        54606.7&  $1.92_{-0.06}^{+0.06}$ & $0.72_{-0.14}^{+0.11}$ & $0.23_{-0.09}^{+0.10}$ & $4.93_{-0.05}^{+0.05}$ &38.5/43 \\
93433-01-08-03 &        54608.7&  $2.05_{-0.03}^{+0.03}$ & $0.05_{-0.04}^{+0.25}$ & $0.01_{-0.01}^{+0.01}$ & $4.21_{-0.05}^{+0.05}$ &44.7/43 \\
93433-01-08-04 &        54606.0&  $1.98_{-0.09}^{+0.07}$ & $0.49_{-0.36}^{+0.25}$ & $0.14_{-0.08}^{+0.14}$ & $4.78_{-0.07}^{+0.07}$ &32.1/43 \\
93433-01-09-00 &        54609.4&  $1.96_{-0.10}^{+0.09}$ & $0.61_{-0.25}^{+0.16}$ & $0.20_{-0.11}^{+0.13}$ & $3.99_{-0.07}^{+0.07}$ &23.2/43 \\
93433-01-09-01 &        54610.4&  $1.92_{-0.10}^{+0.09}$ & $0.79_{-0.14}^{+0.11}$ & $0.29_{-0.13}^{+0.13}$ & $3.86_{-0.05}^{+0.05}$ &26.9/43 \\
93433-01-09-02 &        54611.5&  $1.94_{-0.06}^{+0.06}$ & $0.63_{-0.13}^{+0.11}$ & $0.22_{-0.08}^{+0.09}$ & $4.44_{-0.05}^{+0.05}$ &26.2/43 \\
93433-01-09-03 &        54612.4&  $2.10_{-0.15}^{+0.09}$ & $0.55_{-0.54}^{+0.32}$ & $0.10_{-0.08}^{+0.08}$ & $3.43_{-0.08}^{+0.08}$ &31.9/43 \\
93433-01-09-04 &        54613.6&  $1.98_{-0.10}^{+0.09}$ & $0.61_{-0.24}^{+0.16}$ & $0.19_{-0.11}^{+0.13}$ & $3.81_{-0.07}^{+0.07}$ &34.0/43 \\
93433-01-10-00 &        54618.8&  $2.00_{-0.10}^{+0.09}$ & $0.53_{-0.22}^{+0.16}$ & $0.15_{-0.07}^{+0.09}$ & $3.10_{-0.05}^{+0.05}$ &26.4/43 \\
93433-01-10-02 &        54622.0&  $1.97_{-0.15}^{+0.14}$ & $0.65_{-0.29}^{+0.16}$ & $0.21_{-0.13}^{+0.15}$ & $3.02_{-0.07}^{+0.07}$ &26.0/43 \\
93433-01-11-00 &        54624.8&  $2.01_{-0.12}^{+0.11}$ & $0.67_{-0.23}^{+0.14}$ & $0.17_{-0.10}^{+0.11}$ & $2.76_{-0.05}^{+0.05}$ &35.2/43 \\
93433-01-11-01 &        54628.1&  $2.01_{-0.21}^{+0.20}$ & $0.72_{-0.71}^{+0.21}$ & $0.18_{-0.13}^{+0.11}$ & $2.52_{-0.08}^{+0.08}$ &24.5/43 \\
93433-01-12-00 &        54631.8&  $1.98_{-0.15}^{+0.14}$ & $0.67_{-0.18}^{+0.12}$ & $0.18_{-0.09}^{+0.10}$ & $2.08_{-0.05}^{+0.05}$ &29.9/43 \\
93433-01-12-01 &        54636.7&  $1.99_{-0.17}^{+0.19}$ & $0.70_{-0.56}^{+0.16}$ & $0.17_{-0.08}^{+0.07}$ & $2.32_{-0.06}^{+0.06}$ &34.3/43 \\

\hline
\end{tabular}
\caption{An additional BB component is included in the spectral fitting.
$f_{\rm BB}$: unabsorbed (3--25 keV) flux of BB component in units of
$10^{-10}$ erg cm$^{-2}$ s$^{-1}$. Flux: 3--25 keV unabsorbed flux in units of
$10^{-10}$ erg cm$^{-2}$ s$^{-1}$. All errors are at the 90\% confidence
level.} \label{bb_fits}
\end{table*}

\section{Summary}

Investigating the X-ray archival data, we determined the distance of XTE
J1810-189 as 3.5--8.7 kpc via the PRE burst for the first time. The source
presents some peculiar X-ray behaviours, which distinguish it from most other
NS XRBs: (1) During the 2008 outburst, XTE J1810-189 did not enter into the
high/soft state, and moved steadily from the top right corner to the lower left
corner in the CCD as the accretion rate decreased, as if the X-ray luminosity
were proportional to the accretion rate; (2) The source is highly variable, and
the fractional rms remains at a nearly constant level of $\sim 30$ per cent
with evolving spectra; (3) Both soft and hard colours increase with the
intensity. As the X-ray luminosity decreased from $4\times10^{36}$ ergs
s$^{-1}$ to $6\times10^{35}$ ergs s$^{-1}$, the X-ray spectra became softer
with the photon index $\Gamma$ increasing from $1.84\pm0.01$ to $2.25\pm0.04$.
The dense observations, relatively high flux and broadband spectra allow us to
provide the strong evidence for softening of the non-thermal component (rather
than NS surface emissions as suggested in literature) at low luminosity. Our
results also confirm that NS XRBs have softer spectra below $10^{36}$ erg
s$^{-1}$ than those of BH XRBs, indicating there are different accretion
mechanisms in two classes of XRBs, probably due to the boundary layer existing
in NS XRBs.

\section*{Acknowledgements}

We thank the referee for helpful comments which significantly improved this
work. We are grateful to Long Ji for help discussions on type I X-ray burst
data analysis. This work is partially supported with funding by 973 Program of
China under grant 2014CB845802, the National Natural Science Foundation of
China under grants 11133002, 11373036, and 11303022, the Qianren start-up grant
292012312D1117210, and by the Strategic Priority Research Program ``The
Emergence of Cosmological Structures'' of the Chinese Academy of Sciences,
Grant No. XDB09000000. S.S.W. is funded by the Co-Circulation Scheme, supported
by the EC-FP7 Marie Curie Actions-People-COFUND and T\"{U}B\.{I}TAK.


\end{document}